\documentclass[journal,english,twocolumn]{IEEEtran}

\usepackage[T1]{fontenc}
\usepackage{graphicx}
\usepackage{amssymb}
\usepackage{subfigure}
\usepackage{gensymb}
\usepackage{cite}
\usepackage{siunitx}
\usepackage[american]{circuitikz}
\usetikzlibrary{shapes,arrows}
\usepackage{amsmath,amsfonts,amssymb,amsthm}
\usepackage{mathtools}
\usepackage{commath}


\usepackage{babel}
\addto\captionsenglish{} 

\begin{document}
\bstctlcite{IEEEexample:BSTcontrol}

\title{Design Equations for a Closely-Spaced Two-Element Interferometer Including Internal Noise Coupling}

\author{Adrian T. Sutinjo, Benjamin McKinley, Leonid Belostotski, Daniel C. X. Ung, Jishnu N. Thekkeppattu

\thanks{\textbf{A 4-page letter version of this paper has been accepted for publication by \emph{Radio Science Letters} titled ``Design Equations for Closely-Spaced Two-Element Interferometer for Radio Cosmology'' on 9 Sep. 2020.}}}

\maketitle

\begin{abstract}
We present design equations for a two-element closely-spaced interferometer for measuring the noise temperature of a uniform sky. Such an interferometer is useful for observing highly diffuse radio sources such as the Milky Way and Cosmological signals. We develop a simple equivalent circuit based on radiophysics and antenna theory to describe the interactions between key design parameters such as antenna self and mutual impedance and noise parameters of the receiver; the latter is considered internal noise. This approach straightforwardly facilitates design studies as the response of the uniform signal and the systematic error due to internal noise coupling can be analyzed using the same equivalent circuit. The equivalent circuit shows that mutual coherence due to internal noise coupling is non-negligible and an inherent property of a closely-spaced interferometer. A realistic example design involving two closely-spaced horizontal dipoles over a lossy ground for Cosmological signal detection from 50 to 100\,MHz is discussed as an illustration.
\end{abstract}


\thispagestyle{empty}

\section{Introduction}
\label{sec:intro}
A closely-spaced interferometer provides a detectable response to the isotropic noise temperature component of the surrounding medium. In radio astronomy, the isotropic noise component in question originates from celestial sources located in the far-field of the antennas. In radio interferometry imaging, the close-spacing response is desirable for observation of extended sources~\cite{ekers_rots_1979, Braun_1985A&A...143..307B, 10.1093/mnras/stw2337}. In radio cosmology, the signal of interest is a very small perturbation in the order of 10's to 100's $\si{\milli\kelvin}$ in the Cosmic Microwave Background (CMB) of $\SI{3}{\kelvin}$ in the frequency range of approximately 20 to 200\,MHz~\cite{Pritchard_2012}. This perturbation is due to the emission of 21-cm photons by neutral hydrogen atoms in the early Universe, and the decoupling of the neutral hydrogen spin temperature from the CMB due to various well-understood physical processes~\cite{Madau_1997, FURLANETTO2006181}. This cosmological signal is expected to be rich in spectral structure, containing multiple zero-crossings of higher order derivatives. Observation of this signal must contend with the presence of Galactic noise which is $\sim10^2$ to $\SI{e4}{\kelvin}$ in this frequency range and extra-galactic point-source foregrounds. The foreground signals, however, are expected to be spectrally-smooth, allowing a possible means of separating them from the cosmological signal~\cite{10.1111/j.1365-2966.2011.18276.x}.  A recent report for the EDGES experiment claims the first detection of this signal in the 50 to 100\,MHz range using a single antenna radiometer~\cite{2018Natur.555...67B, Monsalve_2017}. The interest now is to verify this claimed detection. Ideally, a different approach to the measurement should be used in order to avoid similar systematic effects, hence the idea of using a close-spacing interferometer~\cite{Presley_2015, Singh_2015, Mahesh_7265020, Venumadhav_2016}. 

In view of the demand for extreme precision in the measurement and calibration of such a system, the purpose of this paper is to review the fundamental equations that govern closely-spaced antennas such that the key interdependence between critical parameters can be identified and understood. Closely-spaced antennas form a system that can be very highly coupled. The importance of this effect has been recognized in the phased array community~\cite{Warnick_5062509, 7079488, warnick_maaskant_ivashina_davidson_jeffs_2018}, however it has not been incorporated in the analysis of close-spacing interferometers~\cite{Presley_2015, Singh_2015} nor has it translated into a clear design study. Our aim here is a set of design equations that correctly accounts for this complexity while being sufficiently intuitive such that design trends remain easily identifiable. We achieve this with an equivalent circuit and design formulas that do not involve matrix inverses. 

The internal noise generated by the receiver connected to the antennas is radiated by the antenna and picked up by the neighbor. This aspect is particularly critical since the primary argument in favor of the closely-spaced interferometer is that the internal noise does not correlate at the output of the correlator. However as we will demonstrate, the noise coupling contribution is an inherent property of the close-spacing interferometer. This is different from other instrumental effects such as correlated noise due to ohmic losses and cross-talk for which mitigation strategies exist and their limits are well-understood.

The paper is organized as follows. Sec.~\ref{sec:back} and Sec.~\ref{sec:theory} review the context and develop the formulas including the equivalent circuit. Example calculations using the formulas are given in Sec.~\ref{sec:calc}. More realistic examples including considerations for lossy soil are discussed in Sec.~\ref{sec:half-space}.







\section{Background}
\label{sec:back}

\begin{figure}[htb]
    \centering
\begin{circuitikz}[scale = 0.8,transform shape]
\draw 
(0,0) node [antenna](A1){}
(A1.1) to[short,*-,l=$V_{1}$]++(0,-0.5)
(0,-1) node[adder] (add1) {} 
(add1.1) to [short,-*,l_=$V_{i1}$] ++(-0.5,0) 
(add1.2) to [short,l=$V_{L1}$] ++(0,-0.5)
(0,-2) to[amp,*-,t=$g_1$](0,-4);
\draw 
(2,0) node [antenna](A2){}
(A2.1) to[short,*-,l=$V_{2}$]++(0,-0.5)
(2,-1) node[adder] (add2) {} 
(add2.3) to [short,-*,l=$V_{i2}$] ++(0.5,0)
(add2.2) to [short,l=$V_{L2}$] ++(0,-0.5)
(2,-2) to[amp,*-,t=$g_2$](2,-4);
\draw [dashed] (-1.5,0) rectangle
(3.5,2.5)
node[pos=0.5,above]{};
\draw (-1.5,-4) rectangle (3.5,-5)
node[pos=0.5,below]{Correlator};
\end{circuitikz}
    \caption{The block diagram of a two-element interferometer. $V_{i1}, V_{i1}$ are the additive internal noise sources. The observation is made using the correlator output.}
    \label{fig:two_elem_block}
\end{figure}
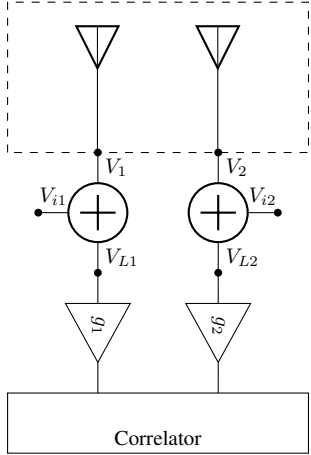

Fig.~\ref{fig:two_elem_block} depicts the two-element interferometer in question. The antennas are exposed to the external environment demarcated by the dashed rectangle. The voltages $V_1, V_2$ are the voltages seen at the antenna ports due to external signals which in our application appear as noise sources. These voltages are extremely faint and require amplification provided by the voltage gain blocks $g_1, g_2$. However, this amplification adds internal noise which we denote as $V_{i1},V_{i2}$.

Unless otherwise noted, we take the voltage and current quantities as complex envelopes, e.g. $V=|V|e^{j\varphi}$ where magnitude $|V|$ is given in rms. The input to the gain blocks is the column vector  $\mathbf{v}_L=[V_{L1}, V_{L2}]^T$ where $V_{L1}=V_1+V_{i1}$ is the external signal contaminated with noise (similarly with $V_{L2}$). The output of the correlator is

\begin{eqnarray}
\mathbf{C}_c=\mathbf{G}\left<\mathbf{v}_L\mathbf{v}_L^H\right>\mathbf{G}^H
=\mathbf{G}\mathbf{C}_L\mathbf{G}^H
\label{eqn:Rc} 
\end{eqnarray}
where $\mathbf{G}$ is a voltage gain matrix which we can initially think of as a diagonal matrix $\text{diag}(g_1, g_2)$ and
\begin{eqnarray}
\mathbf{C}_L&=&
\left[ \begin{array}{cc}
\left<|V_{L1}|^2\right> & \left<V_{L1}V_{L2}^*\right> \\
\left<V_{L1}^*V_{L2}\right> &\left<|V_{L2}|^2\right> \end{array}\right] \nonumber \\
&=&
\left[ \begin{array}{cc}
\left<|V_{1}|^2\right> & \left<V_{1}V_{2}^*\right> \\
\left<V_{1}^*V_{2}\right> &\left<|V_{2}|^2\right> \end{array}\right] +
\left[ \begin{array}{cc}
\left<|V_{i1}|^2\right> & \left<V_{i1}V_{i2}^*\right> \\
\left<V_{i1}^*V_{i2}\right> &\left<|V_{i2}|^2\right> \end{array}\right] \nonumber\\
&=&
\mathbf{C}_a+\mathbf{C}_i
\label{eqn:RL} 
\end{eqnarray}
where the $\mathbf{C}_a$ and $\mathbf{C}_i$ are the coherence matrices of the external noise and the internal noise, respectively. The matrix $\mathbf{C}_L$ is separable as shown in \eqref{eqn:RL} because the internal and external noise sources are uncorrelated. 

\subsection{Single Element Observation}
\label{sec:single}
Equation \eqref{eqn:RL} suggests that a single antenna observes $\left<|V_{1}|^2\right>+\left<|V_{i1}|^2\right>$ which is a quantity proportional to the external noise power added with internal noise power. Hence, to observe the isotropic noise temperature in $\left<|V_{1}|^2\right>$, precise knowledge of internal noise power contribution $\left<|V_{i1}|^2\right>$ is required such that its contribution may be subtracted from the data. 

\subsection{Two-Element Interferometer}
\label{sec:two-elem}
For a two-element interferometer, the off-diagonal contains $\left<V_{1}V_{2}^*\right>+\left<V_{i1}V_{i2}^*\right>$. It is normally \emph{assumed} in current literature in close-spacing interferometer~\cite{Presley_2015, Singh_2015} that $\left<V_{i1}V_{i2}^*\right>=0$ such that only the desired quantity $\left<V_{1}V_{2}^*\right>$ is expected. Furthermore if the spacing is very close, it is argued that $\left<V_{1}V_{2}^*\right> \rightarrow \left<|V_{1}|^2\right>$ for identical antennas. Hence, the basic idea with a two-element interferometer is that it is sensitive to the isotropic temperature while unaffected by the internal noise. These assertions require closer review. To what degree can we expect
\begin{enumerate}
\item $\left<V_{1}V_{2}^*\right> \rightarrow \left<|V_{1}|^2\right>$?
\item $\left<V_{i1}V_{i2}^*\right>=0$?
\end{enumerate}

\section{Theory and Key Equations}
\label{sec:theory}
\subsection{External Noise}
\label{sec:ext noise}
\begin{figure}[htb]
    \centering
\begin{circuitikz}[scale = 0.8,transform shape]
\draw 
(0,0) node[fourport] (c) {$\mathbf{Z}$, noiseless}
(c.4) to[short,i<_=$I_1$] ++(-0.5,0) to [V,invert,-*,l_=$V_{e1}$] (-3,0.45)
(c.3) to[short,i<=$I_2$] ++(0.5,0) to [V,invert,-*,l=$V_{e2}$] (3,0.45)
(c.1) to[short,-*] ++(-2.1,0) 
(c.2) to[short,-*] ++(2.1,0) 
(c.4) node[below right] {$~Z_{11}$}
(c.3) node[below left] {$Z_{12}~$}
(c.1) node[above right] {$~Z_{21}$}
(c.2) node[above left] {$Z_{22}~$}
;
\draw
(-3,0.45) to [open, v_=$V_1$] (-3,-0.5)
(3,0.45) to [open, v^=$V_2$] (3,-0.5);
;
\end{circuitikz}
    \caption{Antenna two-port and noise. The noise voltages $V_{e1}, V_{e2}$ are separated from the noiseless antenna two-port represented by the impedance matrix $\mathbf{Z}$.}
    \label{fig:antenna_noise}
\end{figure}
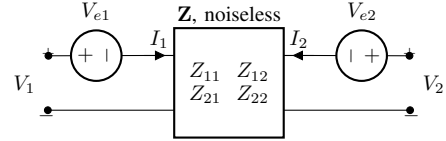

We begin with the first question, i.e. the behaviour of the antenna system and the noise therein as depicted in Fig.~\ref{fig:antenna_noise}. In this representation, the external noise sources $V_{e1},V_{e2}$ are separated from the noiseless antenna system given by the impedance matrix $\mathbf{Z}$~\cite{Rothe_4052096,Hillbrand_1084200}.
\begin{eqnarray}
V_1&=&I_1Z_{11}+I_2Z_{12}+V_{e1} \nonumber \\
V_2&=&I_1Z_{21}+I_2Z_{22}+V_{e2} 
\label{eqn:Zandnoise} 
\end{eqnarray}
Using vector and matrix notations we can write \eqref{eqn:Zandnoise} as
\begin{eqnarray}
\mathbf{v}&=&\mathbf{Z}\mathbf{i}+\mathbf{v}_e
\label{eqn:Zandnoise_vec} 
\end{eqnarray}
The external noise voltages $\mathbf{v}_e=[V_{e1},V_{e2}]^T$ are random such that only the statistics may be quantified.

The mutual coherence of $V_{e1}$ and $V_{e2}$ due to \emph{unpolarized far-field} noise sources characterized by noise temperature $T_{ext}(\theta,\phi)$ in the spherical coordinate system is
\begin{eqnarray}
\frac{\left<V_{e1:ff}V_{e2:ff}^*\right>}{\Delta f}=
\frac{\eta_0 k}{\lambda^2}\int\limits_{0}^{2\pi}\int\limits_{0}^{\pi} T_{ext}(\theta,\phi) [\bar{l}_1\cdot \bar{l}_2^*]\sin\theta d\theta d\phi
\label{eqn:Vecorr} 
\end{eqnarray}
where $\bar{l}_{1,2}=\hat{\theta} l_{1,2}^{\theta} (\theta,\phi)+ \hat{\phi} l_{1,2}^\phi(\theta,\phi)$ is the \emph{open-circuit} effective antenna length (vector), $\eta_0=\sqrt{\mu_0/\epsilon_0}\approx120\pi\,\Omega$ is the free space impedance, $k$ is the Boltzmann constant and $\lambda$ is the wavelength at the center frequency of the observation/calculation and $\Delta f$ is the frequency bandwidth of that observation. The open-circuit antenna effective lengths $\bar{l}_1,\bar{l}_2$ are taken with respect to a common coordinate origin $(0,0)$ in the presence of the other element which is open-circuited at its port. The full detail of the external noise as given by~\eqref{eqn:Vecorr} is important for foreground removal and detection for which there is a significant body of work as reviewed in the introduction. Our work here, however, focuses on instrumental effects not typically considered in that literature and therefore we consider the following specializations.

In the special case of isotropic noise temperature $T_{iso}$
\begin{eqnarray}
\frac{\left<V_{e1:ff_{iso}}V_{e2:ff_{iso}}^*\right>}{\Delta f}=
\frac{\eta_0 k}{\lambda^2}T_{iso}\int\limits_{0}^{2\pi}\int\limits_{0}^{\pi}  \bar{l}_1\cdot \bar{l}_2^*\sin\theta d\theta d\phi
\label{eqn:Vecorr_iso} 
\end{eqnarray}
Once the antennas are decided, equations \eqref{eqn:Vecorr}, \eqref{eqn:Vecorr_iso} may be easily computed. However these equations are not convenient for design because $\bar{l}_{1,2}$ are complex vector quantities that vary over $\theta,\phi$. Furthermore, the integral of the inner product over the sphere is not intuitive. We need a meaningful simplification. 

\emph{Special case}: If all external noise sources are at a thermal equilibrium\footnote[1]{This can be achieved if the radiation efficiencies of both antennas are 100\% or if the ohmic losses are at the same temperature as $T_{iso}$. The latter does not generally apply to sky observation, however, it applies to measurements of the said system in an anechoic chamber at ambient temperature, $T_{iso}=T_{amb}$.}  at temperature $T_{iso}$ then the coherence matrix of the external noise voltages is referred to as the generalized Nyquist theorem~\cite{doi:10.1063/1.1722048, Hillbrand_1084200} 
\begin{eqnarray}
\mathbf{C}_e|_{T_{iso}}=\left<\mathbf{v}_e\mathbf{v}_e^H\right>|_{T_{iso}}=4kT_{iso}\Delta f\Re [\mathbf{Z}] 
\label{eqn:Re} 
\end{eqnarray}
where
\begin{eqnarray}
\Re [\mathbf{Z}] = \left[ \begin{array}{cc}
R_{11}& R_{12} \\
R_{21} & R_{22} \end{array}\right]
\label{eqn:ReZ} 
\end{eqnarray}
These are antenna two-port quantities, each of which is a single real number at every frequency. The quantity of interest for a two-element interferometer is 
\begin{eqnarray}
\left<V_{e1:T_{iso}}V_{e2:T_{iso}}^*\right>=4kT_{iso}\Delta f R_{12}
\label{eqn:Vecorr_Tiso} 
\end{eqnarray}
where $R_{12}$ is the antenna mutual resistance. This is a standard quantity that has been studied in the antenna community for many decades. The results have been tabulated and plotted~\cite{Brown_1685626, Kraus_ch10, Alex_1142531, Stutz_ch8} and is now easily computed using electromagnetic simulation as well as measured using a standard test equipment such as a vector network analyzer (VNA). The single antenna response is
\begin{eqnarray}
\left<|V_{e1:T_{iso}}|^2\right>=4kT_{iso}\Delta f R_{11}
\label{eqn:Vesingle_Tiso} 
\end{eqnarray}
Therefore if we are somehow able to detect $V_1$ and $V_2$ without incurring impedance loading effect nor additive noise, then the single element response is $\propto R_{11}$ while the two-element interferometer response is $\propto R_{12}$. 

\begin{figure}[htb]
\centering
\noindent
  \includegraphics[width=2.75in]{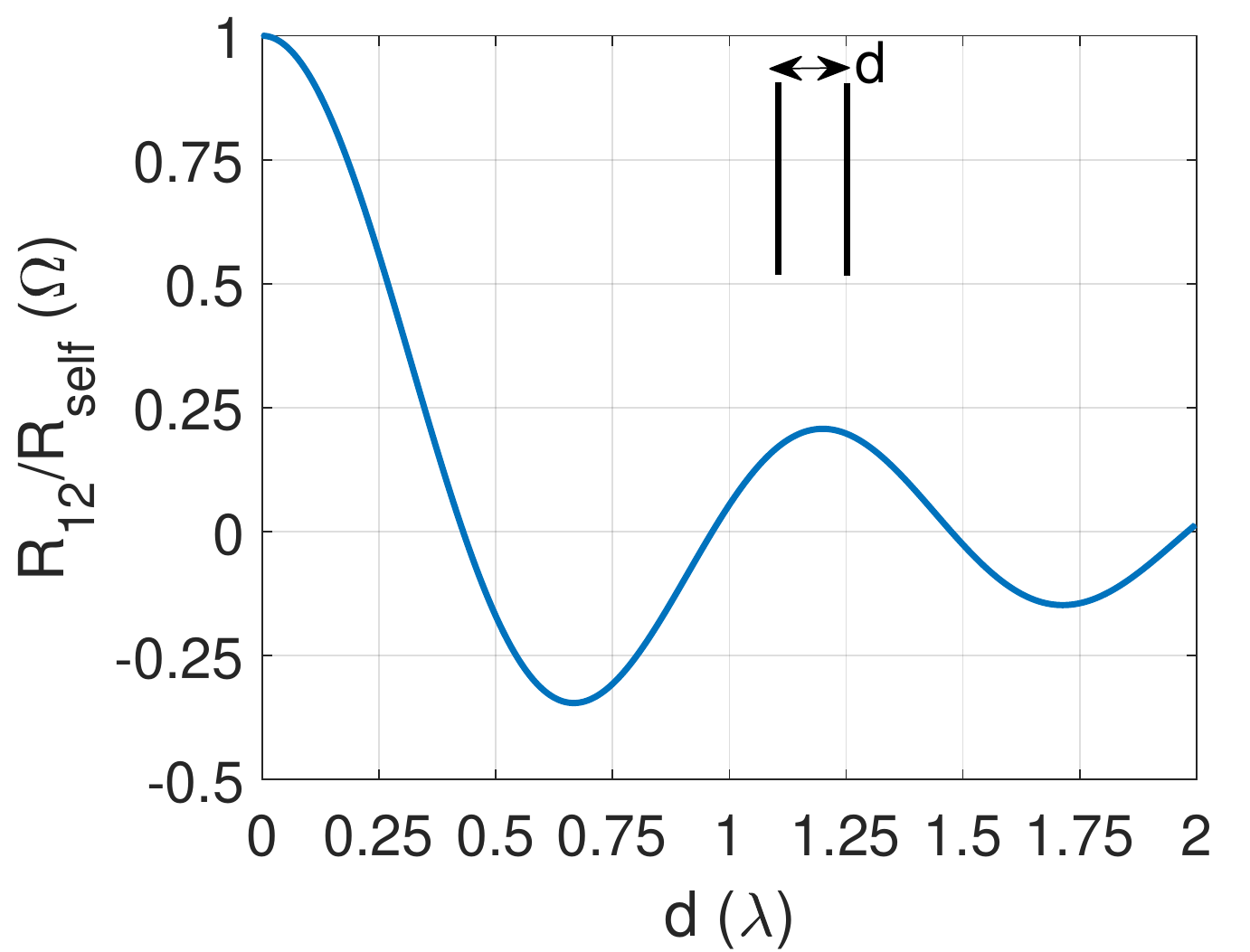}
\caption{Ratio of mutual resistance to self resistance of parallel thin dipoles $0.48\lambda$ long vs. distance, $d$. In this case $R_{self}\approx R_{11}$. The zero crossings in $\lambda$ are $0.43, 0.96, 1.47, 1.98$.}
\label{fig:R21_self}
\end{figure}

It is instructive to review the $R_{11}$ and $R_{12}$ of a simple antenna such as a dipole or a monopole.  Fig.~\ref{fig:R21_self} shows the mutual-resistance-to-self-resistance ratio vs. distance of parallel identical thin $0.48\lambda$ dipoles (just below first resonance) computed using the formulas in~\cite{Brown_1685626, Kraus_ch10}. The plot can also be read as the same quantity for monopoles which are half the dipole lengths. For dipoles with lengths $\lesssim 0.5\lambda$, the $R_{11}\approx R_{self}$ because we can assume that there is negligible current excited in the open-circuited neighboring dipole.

We see that $R_{12}$ converges to $R_{11}$ as $d\rightarrow 0$. This is fully expected and consistent with \eqref{eqn:Vecorr_iso}. However $R_{12}/R_{11}$ drops rapidly with increasing $d$. The curve crosses the first zero at approximately $0.43\lambda$. Zero-crossings are generally undesirable as they represent points at which the two-element interferometer has no response to the isotropic noise. For a wideband two-element interferometer system, the zero crossing points constitute the bandwidth limit of the system. For example, if we place the center frequency of the interferometer at the second peak at $d\approx 1.2\lambda$, then the bandwidth will be limited by the ratio of the zeros crossing points of $1.47\lambda$ and $0.96\lambda$ flanking that point, which results in bandwidth of $1.47/0.96\approx1.5:1$. Fig.~\ref{fig:R21_self} suggests, therefore, that the widest bandwidth is expected by placing the highest frequency below the first zero-crossing of $0.43\lambda$. If we desire a response that is at least half the single element response, then the spacing should be $\lesssim 0.27\lambda$. 

Returning to question 1 posed in Sec.~\ref{sec:two-elem}, this analysis proves quite definitive. We confirm that indeed $\lim_{d\rightarrow0} \left<V_1V_2^*\right>=\left<|V_1|^2\right>$. However, this is not a monotonic behavior and exhibits zero crossings that are bandwidth-limiting. The least bandwidth-limited design space exists for very close spacings $d\lesssim0.43\lambda$ below the first zero-crossing. 

It is also useful to list known trends in mutual impedance to guide design choice~\cite{Stutz_ch8}:
\begin{itemize}
    \item $|Z_{12}|$ decreases with $d$, in many cases $\propto d^{-2}$.
    \item Antennas with sharper beams are less coupled than those with broader beams.
    \item Parallel orientation couple more than collinear orientation.
    \item Larger antennas are less coupled.
\end{itemize}
This list suggests closely-spaced parallel small antennas are strong candidates, which is consistent with the plot in Fig.~\ref{fig:R21_self}. This is also consistent with the findings in~\cite{Singh_2015} which are based only on the Fourier footprint of the antenna beams.

\subsection{Gain Block Input Impedance Loading Effect}
\label{sec:gain_block_ZL}
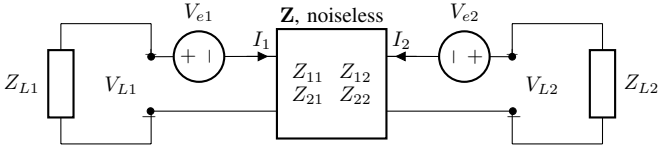
\begin{figure}[htb]
    \centering
\begin{circuitikz}[scale = 0.8,transform shape]
\draw 
(0,0) node[fourport] (c) {$\mathbf{Z}$, noiseless}
(c.4) to[short,i<_=$I_1$] ++(-0.5,0) to [V,invert,-*,l_=$V_{e1}$] (-3,0.45)
(c.3) to[short,i<=$I_2$] ++(0.5,0) to [V,invert,-*,l=$V_{e2}$] (3,0.45)
(c.1) to[short,-*] ++(-2.1,0) 
(c.2) to[short,-*] ++(2.1,0) 
(c.4) node[below right] {$~Z_{11}$}
(c.3) node[below left] {$Z_{12}~$}
(c.1) node[above right] {$~Z_{21}$}
(c.2) node[above left] {$Z_{22}~$}
;
\draw
(-3,0.45) to [open, v_=$V_{L1}$] (-3,-0.5)
(3,0.45) to [open, v^=$V_{L2}$] (3,-0.5);
;
\draw
(-3,0.45) to[short,-] (-3, 1)
to[short,-] (-4.5, 1)
to[generic,l_=$Z_{L1}$] (-4.5,-1)
to[short,-] (-3, -1)
to[short,-] (-3, -0.45)
;
\draw
(3,0.45) to[short,-] (3, 1)
to[short,-] (4.5, 1)
to[generic,l=$Z_{L2}$] (4.5,-1)
to[short,-] (3, -1)
to[short,-] (3, -0.45)
;
\end{circuitikz}
    \caption{Antenna two-port and noise including gain block impedance loading.}
    \label{fig:antenna_noise_with_load}
\end{figure}

The antenna voltages $V_1,V_2$ in Fig.~\ref{fig:antenna_noise} must be detected through additional instrumentation. As suggested in Fig.~\ref{fig:two_elem_block}, we connect gain blocks 1 and 2 to the antenna ports. These gain blocks present finite load impedances $Z_{L1},Z_{L2}$ to ports 1, 2 of the antennas as shown in Fig.~\ref{fig:antenna_noise_with_load}. The voltage seen at the load impedance $Z_{L2}$ is
\begin{eqnarray}
V_{L2}|_{V_{e1}}&=&\frac{-V_{e1}}{Z_{L1}+Z_{11}}Z_{21}\frac{Z_{L2}}{Z_{L2}+Z_{emb2}}\nonumber \\
V_{L2}|_{V_{e2}}&=&V_{e2}\frac{Z_{L2}}{Z_{L2}+Z_{emb2}}\nonumber \\
V_{L2}|_{V_{e1,2}}&=&\frac{Z_{L2}}{Z_{L2}+Z_{emb2}}\left( V_{e2}-\frac{V_{e1}Z_{21}}{Z_{L1}+Z_{11}}\right)
\label{eqn:VL_Ve} 
\end{eqnarray}
The rationale is as follows. Taking $V_{L2}|_{V_{e1}}$ for example, the first factor is the current entering port 1 due to $V_{e1}$; multiplying this factor with $Z_{21}$ produces the Th\'evenin equivalent voltage looking into port 2 ($V_{th2}$) which is in series with the Th\'evenin equivalent impedance which is the embedded antenna impedance $Z_{emb2}$ (the impedance seen at the port of antenna 2 given that port 1 is terminated with $Z_{L1}$). The final factor is a voltage division ratio which transfers the Th\'evenin equivalent voltage to $V_{L2}$. $V_{L1}|_{V_{e1,2}}$ can be found simply by swapping subscripts $._1$ with $._2$ in \eqref{eqn:VL_Ve}.  

The desired quantity is the coherence matrix of the load voltages due to the external noise sources, in particular $\left<V_{L1}V_{L2}^*\right>|_{V_{e1,2}}$. However, in the general case this expression is lengthy and not particularly insightful. For design equations, it is adequate to consider the case where the \emph{antennas are reciprocal} $Z_{21}=Z_{12}$ and \emph{the antennas and the gain blocks are identical} ($Z_{L1}=Z_{L2}=Z_{L}; Z_{11}=Z_{22}, Z_{emb1}=Z_{emb2}=Z_{emb}$). With this assumption and using the conditions that apply for the coherence matrix in \eqref{eqn:Re}, we obtain 
\begin{eqnarray}
\left<V_{L1}V_{L2}^*\right>|_{V_{e1,2}}&=& \left|\frac{Z_{L}}{Z_{L}+Z_{emb}}\right|^2 \left<V_{th1}V_{th2}^*\right>|_{V_{e1,2}}
\label{eqn:ext_noise_corr_loaded} 
\end{eqnarray}
where
\begin{eqnarray}
\frac{\left<V_{th1}V_{th2}^*\right>|_{V_{e1,2}}}{4kT_{iso}\Delta f}&=&
-2\Re\left[\frac{Z_{21}}{Z_{11}+Z_{L}}\right]R_{11}+\cdots \nonumber \\
&+&R_{21}\left(1+\left|\frac{Z_{21}}{Z_{11}+Z_{L}}\right|^2\right)
\label{eqn:ext_noise_corr_loaded_B} 
\end{eqnarray}

\begin{figure*}[htb]
\begin{circuitikz}[scale=1,transform shape]
\draw 
(0,0) to[generic,*-*,l=$Z_{L1}$,v<=$V_{L1}$] (0,3)
to[short] (1.5,3)
to[I, *-*,l=$I_{n1}$] (1.5,0)
to[short] (0,0);
\draw 
(1.5,3)to[short](2.0,3)
to[V,invert,l=$V_{n1}$] (3.0,3)
to[V,-*,l=$V_{e1}$] (4.5,3)
to [open, v^=$V_1'$] (4.5,0) 
to [short,*-] (1.5,0) 
;
\draw 
(4.5,3)to[short, i=$I_1$] (5.5,3)
to[generic,l=$Z_{11}-Z_{21}$] (7.15,3)
to[generic,l_=$Z_{21}$] (7.15,0)
to[short, -*] (4.5,0)
;
\draw
(7,3) to[generic,l=$Z_{22}-Z_{21}$] (9,3)
to[short,-*,i<=$I_2$] (9.5,3)
to [open, v_=$V_2'$] (9.5,0) 
to[short] (7.15,0)
;
\draw 
(9.5,3)to[short](10,3)
to[V,invert,l=$V_{e2}$] (11,3)
to[V,-*,l=$V_{n2}$] (12.5,3)
to[I, *-*,l_=$I_{n2}$] (12.5,0)
to[short,-*] (9.5,0)
;
\draw 
(12.5,3) to[short] (14,3)
(14,3) to[generic,*-*,l=$Z_{L2}$,v=$V_{L2}$] (14,0)
to[short] (12.5,0)
;
\end{circuitikz}
\caption{Equivalent circuit for a two-element interferometer including external ($V_{e1},V_{e2}$) and internal noise sources ($V_{n1},V_{n2}, I_{n1},I_{n2}$). The antennas are reciprocal $Z_{21}=Z_{12}$. In this case, $Z_{emb1}=Z_{11}-Z_{21}+Z_{21}||(Z_{22}-Z_{21}+Z_{L2})$, similarly with $Z_{emb2}$.} \label{fig:equiv_ckt}
\end{figure*}
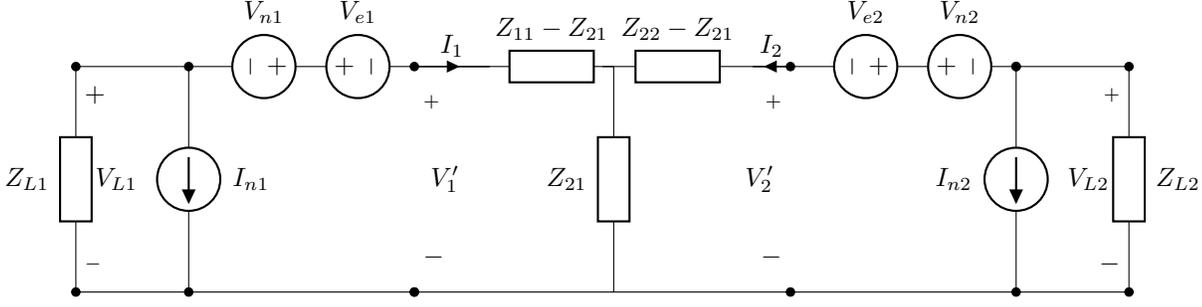

For reciprocal antennas ($Z_{21}=Z_{12}$), the $\mathbf{Z}$ matrix may be replaced with a T~network shown in Fig.~\ref{fig:equiv_ckt} (for now we ignore the internal voltage and current noise sources $V_{i1,2},I_{i1,2}$ by replacing them with short and open circuits, respectively).  This schematic is helpful in interpreting \eqref{eqn:ext_noise_corr_loaded}.  First  we see in \eqref{eqn:ext_noise_corr_loaded} that if $Z_L\rightarrow\infty$, we recover \eqref{eqn:Vecorr_Tiso} as expected. In Fig.~\ref{fig:equiv_ckt}, we also see that exchange of information from the two halves of the circuit is facilitated by voltage division through the T~network. For example, a voltage source on the left (e.g., $V_{e1}$) will appear as a Th\'evenin equivalent voltage source looking into port 2 with voltage division ratio $Z_{21}/(Z_{11}+Z_{L})$ which appears in \eqref{eqn:ext_noise_corr_loaded_B}. This suggests that the exchange of information from each half of the circuit may be \emph{minimized} by making $|Z_{11}+Z_L|\gg |Z_{21}|$; of course in the process, $R_{21}$ must \emph{not} become vanishingly small. This condition is achievable in practice because $|Z_{21}|$ tends to decrease with respect to $|Z_{11}|$ with increasing distance; furthermore at the target VHF frequencies, $Z_L$ can be made high, for example using a field effect transistor (FET) with a high input impedance.       

\subsection{Internal Noise}
\label{sec:int noise}

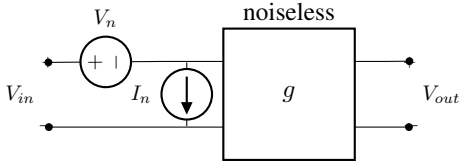
\begin{figure}[htb]
    \centering
\begin{circuitikz}[scale = 0.8,transform shape]
\draw 
(0,0) node[fourport,t=$g$][scale = 1.2] (c) {noiseless}
(c.4) to[short] ++(-1,0) to [V,invert,-*,l_=$V_{n}$] (-4,0.53)
(c.3) to[short] ++(0.5,0) to [short,-*] (2,0.55)
(c.1) to[short,-*] ++(-2.9,0) 
(c.2) to[short,-*] ++(0.9,0) 
;
\draw
(-4,0.5) to [open, v_=$V_{in}$] (-4,-0.5)
(2,0.5) to [open, v^=$V_{out}$] (2,-0.5);
;
\draw
(-1.7,0.53) to[I, -,l_=$I_{n}$] (-1.7,-0.53)
;
\end{circuitikz}
    \caption{Representation of a gain block including the internal noise sources.}
    \label{fig:internal_noise}
\end{figure}

A noisy amplifier may be represented by a partially correlated voltage and noise sources at the input connected to a noiseless two port gain stage~\cite{Rothe_4052096,Haus_4065518, Hillbrand_1084200} as shown in Fig.~\ref{fig:internal_noise}. This information is usually given in data sheets as four noise parameters: $R_n$ (noise resistance, $\si{\ohm}$), $F_{min}$ (minimum noise factor, linear), $Y_{opt}$ (optimum source admittance, $\si{\mho}$) which are convertible as follows~\cite{Hillbrand_1084200}.
\begin{eqnarray}
\frac{\left<|V_{n}|^2\right>}{4kT_0\Delta f}&=&R_n \nonumber \\
\frac{\left<V_nI_n^* \right>}{4kT_0\Delta f}&=&\frac{F_{min}-1}{2}-R_nY_{opt}^* \nonumber \\
\frac{\left<|I_n|^2\right>}{4kT_0\Delta f}&=&R_n|Y_{opt}|^2 
\label{eqn:VI_noise_par}
\end{eqnarray}
where $T_0=\SI{290}{\kelvin}$ is the reference temperature.

With this information, we can complete the equivalent circuit as shown Fig.~\ref{fig:equiv_ckt}. We can now use the circuit to compute $V_{L1,2}$ due to the internal noise source and find the mutual coherence of that voltage. We begin by computing contributions of $V_{n1},I_{n1}$ to $V_{L1},V_{L2}$.
\begin{eqnarray}
V_{L1}|_{V,I_{n1}}&=&\frac{Z_{L1}(-V_{n1}-I_{n1}Z_{emb1})}{Z_{L1}+Z_{emb1}}\nonumber \\
V_{L2}|_{V,I_{n1}}&=&\frac{Z_{21}(V_{n1}-I_{n1}Z_{L1})}{Z_{11}+Z_{L1}}\frac{Z_{L2}}{Z_{L2}+Z_{emb2}}
\label{eqn:Int_noise_vL}
\end{eqnarray}
where the shorthand notation $V,I_{n1}$ has been introduced to represent $V_{n1},I_{n1}$. Assuming identical antennas and gain blocks, 
\begin{eqnarray}
\left<V_{L1}V_{L2}^*\right>|_{V,I_{n1}}&=&\left|\frac{Z_L}{Z_L+Z_{emb}}\right|^2 \left(\frac{Z_{21}}{Z_{11}+Z_L}\right)^*VV|_{V,I_{n1}}\nonumber \\
VV|_{V,I_{n1}}&=&-\left<|V_{n1}|^2\right> +\left<|I_{n1}|^2\right>Z_{emb}Z_L^* +\cdots \nonumber \\
&&+\left<V_{n1}I_{n1}^*\right>Z_L^*-\left<V_{n1}^*I_{n1}\right>Z_{emb}
\label{eqn:Int_noise_vL_corr}
\end{eqnarray}
where $VV|_{V,I_{n1}}$ is a complex voltage times voltage product with units \si{\volt\squared} which has been introduced for brevity and is not to be confused with $V,I_{n1}$. Assuming reciprocal antennas,
\begin{eqnarray}
\left<V_{L1}V_{L2}^*\right>|_{V,I_{n2}}&=&\left|\frac{Z_L}{Z_L+Z_{emb}}\right|^2 \frac{Z_{21}}{Z_{11}+Z_L} VV|_{V,I_{n2}}\nonumber \\
VV|_{V,I_{n2}}&=&-\left<|V_{n2}|^2\right> +\left<|I_{n2}|^2\right>Z_{emb}^*Z_L +\cdots \nonumber \\
&&+\left<V_{n2}^*I_{n2}\right>Z_L-\left<V_{n2}I_{n2}^*\right>Z_{emb}^*
\label{eqn:Int_noise_vL_corr2}
\end{eqnarray}
The subscripts $._1,._2$ may be dropped from the internal noise statistics in \eqref{eqn:Int_noise_vL_corr}, \eqref{eqn:Int_noise_vL_corr2} because the gain blocks are identical such that $(VV|_{V,I_{n1}})^*=VV|_{V,I_{n2}}$. Also, the internal noise sources in gain block 1 are uncorrelated to those of gain block 2; therefore the mutual coherence of antenna load voltages due to internal noise is the sum of \eqref{eqn:Int_noise_vL_corr} and \eqref{eqn:Int_noise_vL_corr2}
\begin{eqnarray}
\left<V_{L1}V_{L2}^*\right>|_{int.}&=&\left<V_{L1}V_{L2}^*\right>|_{V,I_{n1}}+\left<V_{L1}V_{L2}^*\right>|_{V,I_{n2}} \nonumber \\
&=&\left|\frac{Z_L}{Z_L+Z_{emb}}\right|^2 2\Re\left(\frac{Z_{21}VV|_{V,I_{n2}}}{Z_{11}+Z_L}\right) \nonumber \\
VV|_{V,I_{n2}}&=&-\left<|V_{n}|^2\right> +\left<|I_{n}|^2\right>Z_{emb}^*Z_L +\cdots \nonumber \\
&&+\left<V_{n}^*I_{n}\right>Z_L-\left<V_{n}I_{n}^*\right>Z_{emb}^*
\label{eqn:Int_noise_vL_tot}
\end{eqnarray}
Again we see that the leakage of internal noise from the left half of the circuit to the right half and vice versa is through the voltage division $Z_{21}/(Z_{11}+Z_L)$. Therefore, we expect $|Z_{11}+Z_L|\gg |Z_{21}|$ to reduce the mutual coherence due to internal noise coupling. However \eqref{eqn:Int_noise_vL_tot} also shows that $Z_L$ appears in two terms in $VV|_{V,I_{n2}}$. Therefore, we should not be overly reliant on making $Z_L$ large. This is similarly the case with $Z_{11}$ because for large $Z_L$, $Z_{emb}$ approaches $Z_{11}$. Therefore, $|Z_{11}+Z_L|\gg |Z_{21}|$ may be of limited effectiveness to minimize internal noise coherence.

Mutual coherence due to internal noise sources is zero when $2\Re\left[Z_{21}(VV|_{V,I_{n2}})/(Z_{11}+Z_L)\right]=0$. To identify design choices, we study the possibilities to achieve  $VV|_{V,I_{n2}}=0$. We can use the conversion formulas \eqref{eqn:VI_noise_par} in \eqref{eqn:Int_noise_vL_tot}.
\begin{eqnarray}
\frac{VV|_{V,I_{n2}}}{4kT_0\Delta f}&=&-R_n +R_n|Y_{opt}|^2Z_{emb}^*Z_L +\cdots \nonumber \\
&+&\left(\frac{F_{min}-1}{2}-R_nY_{opt} \right)Z_L+\cdots \nonumber\\
&-&\left(\frac{F_{min}-1}{2}-R_nY_{opt}^* \right)Z_{emb}^*
\label{eqn:plug_noise_par}
\end{eqnarray}
After simplification and using $Z_{opt}=1/Y_{opt}$
\begin{eqnarray}
\frac{VV|_{V,I_{n2}}}{4kT_0\Delta f}
&=&-R_n[(1-Z_{emb}^*/Z_{opt}^*)(1+Z_L/Z_{opt})+\cdots \nonumber\\ &+&\frac{F_{min}-1}{2R_n}(Z_{emb}^*-Z_L) ]
\label{eqn:plug_noise_par_simp}
\end{eqnarray}
This suggests possible ways to achieve $VV|_{V,I_{n2}}=0$. For example, if $Z_{emb}=Z_{opt}$, then the first term in the square bracket in \eqref{eqn:plug_noise_par_simp} vanishes and we are left with
\begin{eqnarray}
\frac{VV|_{V,I_{n2}}}{4kT_0\Delta f}\Big|_{Z_{emb}=Z_{opt}}=\frac{F_{min}-1}{2}(Z_L-Z_{emb}^*)
\label{eqn:VVn_Zemb_Zopt}
\end{eqnarray}
Because $F_{min}>1$, \eqref{eqn:VVn_Zemb_Zopt} can only be made zero by conjugate matching $Z_L=Z_{emb}^*$. This means one way we can achieve zero internal noise coherence is $Z_L^*=Z_{emb}=Z_{opt}$.
\begin{eqnarray}
\frac{VV|_{V,I_{n2}}}{4kT_0\Delta f}= 0~\text{for}~Z_L^*=Z_{emb}=Z_{opt}
\label{eqn:zero_noise_coherence}
\end{eqnarray}
A special case that approaches \eqref{eqn:zero_noise_coherence} is $Y_{opt}=0 ~(Z_{opt}=\infty)$ such that the current noise source $\left<|I_n|^2\right>=0$ which suggests an LNA that only has a voltage noise source. In this case, no noise coupling occurs if $Z_L^*=\infty$ since no current due to $V_{n1,2}$ enters the T junction. The open circuit load impedance can only be approached, but not completely fulfilled, in practice with a FET LNA. Hence, to remove noise coupling we need $Z_{emb}\rightarrow~\infty$ which is fulfilled only for a vanishingly small dipole antenna for which $R_{21}$ also vanishes. This special case illustrates the inherent property of internal noise coupling in a closely-spaced interferometer since the same $R_{21}$ that produces the desired external response also couples the undesired internal noise. Returning to the second question posed in Sec.~\ref{sec:two-elem}, we conclude that the mutual coherence due to internal noise does \emph{not} generally vanish. On the contrary, internal noise coupling is inherent to closely-space interferometers. 

\subsection{Summary}
\label{sec:summary}
In summary to the questions posed in Sec.~\ref{sec:two-elem}, based on the analyses in this section, we can say:
\begin{enumerate}
\item $\left<V_{1}V_{2}^*\right> \rightarrow \left<|V_{1}|^2\right>$? Yes, this is fulfilled as spacing approaches zero. It is approximately true for sub-wavelength spacings, see Fig.~\ref{fig:R21_self}.
\item $\left<V_{i1}V_{i2}^*\right>=0$? No, generally this is \emph{not} true; e.g., see \eqref{eqn:Int_noise_vL_tot},  \eqref{eqn:plug_noise_par_simp} and the discussion below \eqref{eqn:zero_noise_coherence}. 
\end{enumerate}

\section{Example Calculations}
\label{sec:calc}
\subsection{Single Dipole}
\label{sec:sing_dip}
\begin{figure}[htb]
\centering
\noindent
  \includegraphics[width=2.75in]{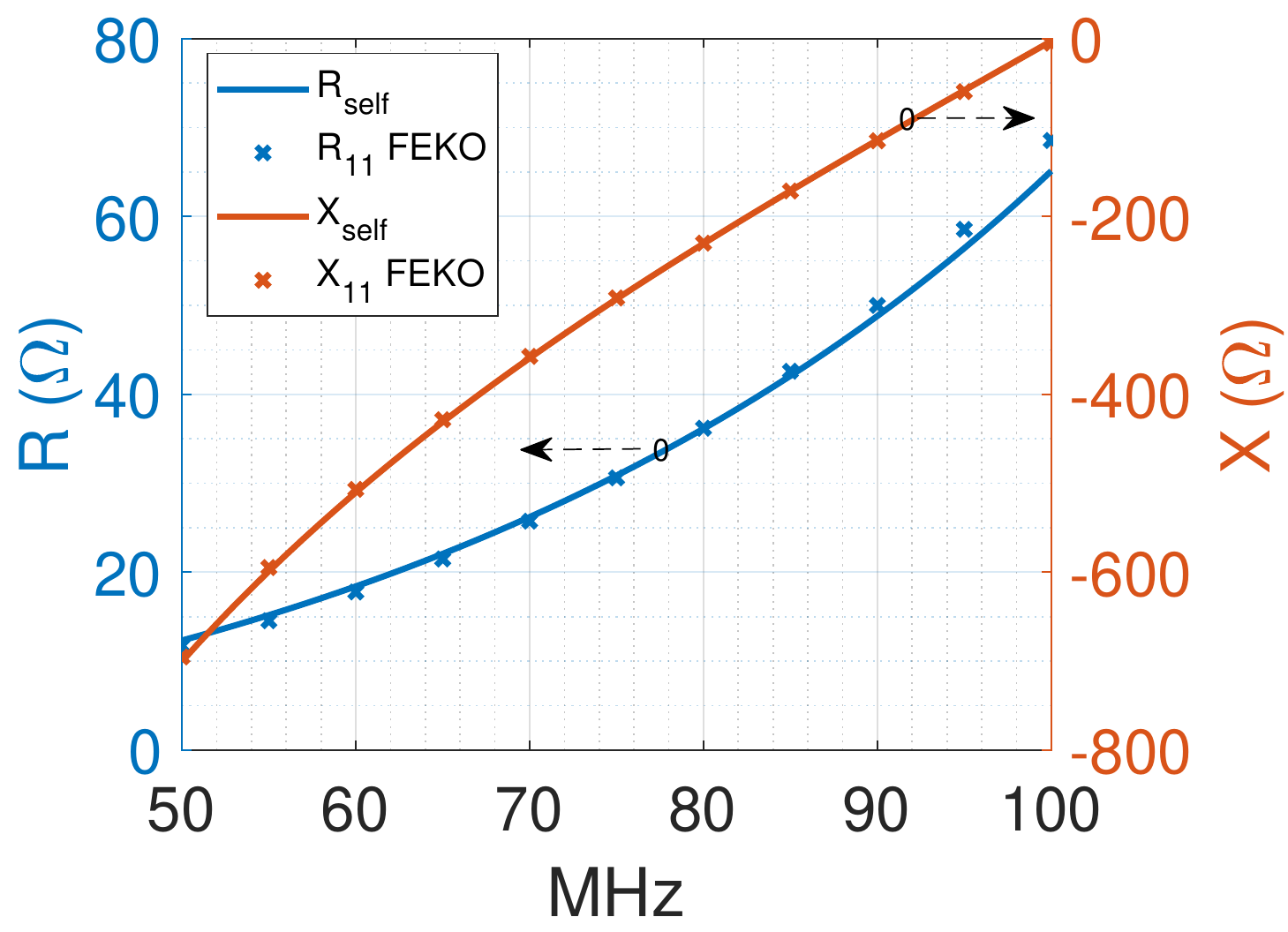}
\caption{Calculated and simulated self impedance of a thin dipole $L=\SI{1.44}{\metre}$ with wire radius of \SI{1}{\milli\metre}.}
\label{fig:Z_self_144}
\end{figure}

As a basis for comparison, consider a 50 to 100\,MHz single antenna system. We select a dipole with length $L=\SI{1.44}{\metre}$ ($0.48\lambda$ at 100\,MHz). The self impedance plotted in Fig.~\ref{fig:Z_self_144} is calculated using the formula in~\cite{Brown_1685626} which compares very well to the simulated results using the Method-of-Moments (MoM) FEKO package with vertex feeding.  The antenna is just below series resonance at 100\,MHz. 

We connect this antenna to a pHEMT low noise amplifier (LNA) Minicircuits SAV-541+\footnote[2]{https://www.minicircuits.com/pdfs/SAV-541+.pdf} which is marketed as an ultra low-noise amplifier. We select the bias point of $V_{DS}=\SI{3}{\volt}, I_{D}=\SI{60}{\milli\ampere}$ as an example. The lowest available $S_{11}$ data point is 100\,MHz. By plotting and averaging the lowest few data points, we infer that the input of the LNA may be modelled as \SI{11}{\ohm} resistor in series with a \SI{5}{\pico\farad} capacitor.  The noise parameters are listed only down to 500\,MHz. To extrapolate  to lower frequencies, we examine $F_{min}$, $G_{opt}$ and $R_n$ over the 3 lowest frequency points in the data sheet (500, 700, 900\,MHz). The clearest trends are $T_{min}=T_0(F_{min}-1) \propto f$, $|B_{opt}|\propto f$ which agree with pHEMT noise models~\cite{Cappy_3475, Pospi_32217}, whereas $G_{opt}$ is nearly constant and $R_n/50$ has a slight slope of $-\num{3e-3}/\SI{100}{\mega\hertz}$. As first-order approximation based on the data sheet values
\begin{eqnarray}
F_{min}&=&1.025 \nonumber \\
R_n/50&=&0.05 \nonumber \\
G_{opt}&=&\SI{0.0106}{\mho} \nonumber \\
B_{opt}&=&-\SI{0.0017}{\mho}\times\frac{f_{MHz}}{500} 
\label{eqn:noise_par}
\end{eqnarray}
The $F_{min}$ has been kept constant because $T_{min}\propto f$ does not hold indefinitely for a realistic LNA due to $1/f$ noise and lossy input bias circuit which eventually limit the $F_{min}$ performance at low frequencies.  

For a single element, the input-referred noise temperature of the receiver is given by~\cite{Lange_1049782}
\begin{eqnarray}
T_{single}&=&T_0(F_{single}-1) \nonumber \\
F_{single}&=&F_{min}+\frac{R_n}{\Re[Y_{self}]}|Y_{self}-Y_{opt}|^2
\label{eqn:T_single}
\end{eqnarray}

\begin{figure}[htb]
\centering
\noindent
  \includegraphics[width=2.75in]{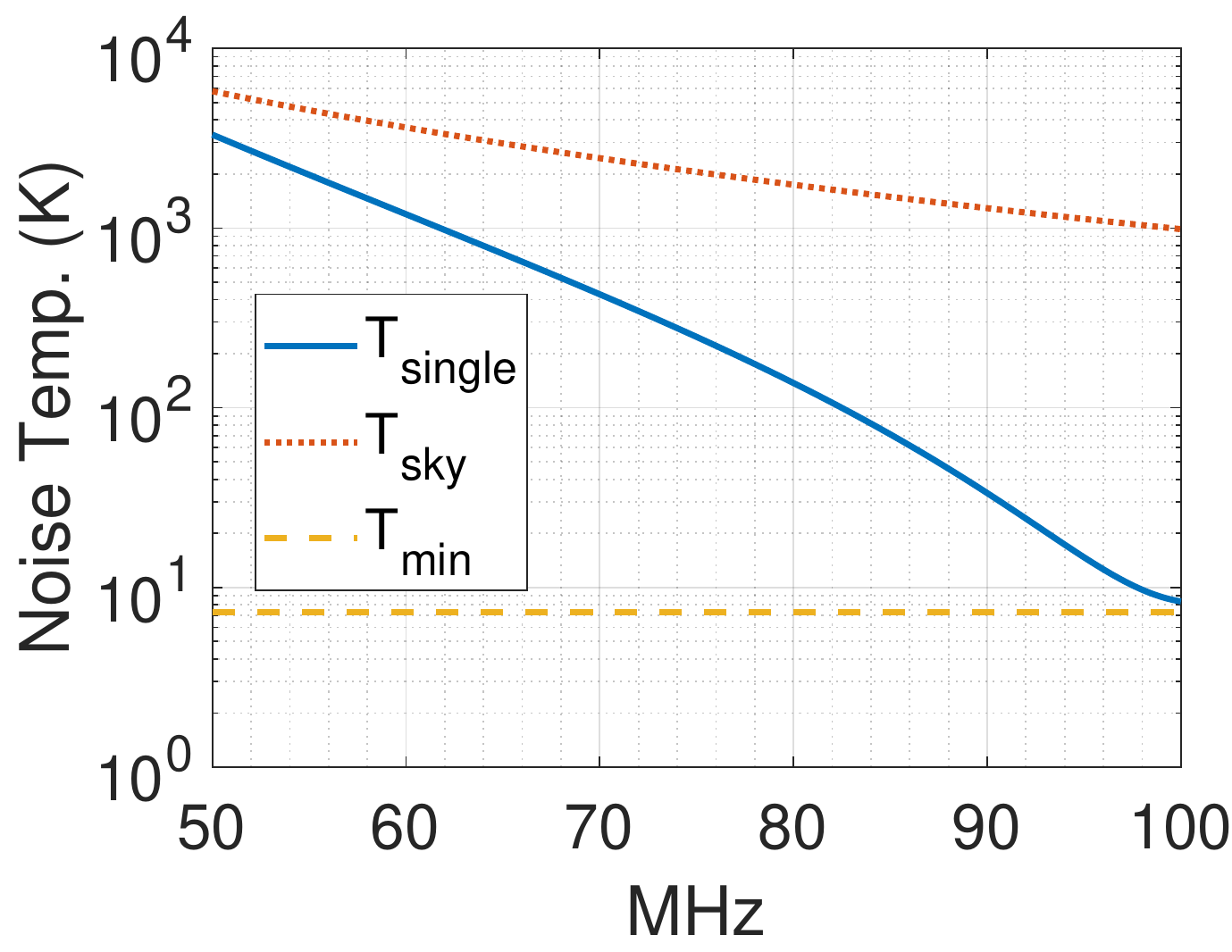}
\caption{Input-referred noise temperatures of the sky, single antenna (thin dipole $L=\SI{1.44}{\metre}$) and the minimum noise temperature of the LNA.}
\label{fig:T_single}
\end{figure}

We plot $T_{single}$ for this example in Fig.~\ref{fig:T_single}. For comparison we include an average sky temperature, $T_{sky}=60\lambda^{2.55}\,\si{\kelvin}$~\cite{nijboer2013lofar}, which is primarily due to the Galactic noise; also included is the lowest achievable noise temperature $T_{min}$ which is realized when $Y_{opt}=Y_{self}$. At 50\,MHz, $T_{single}$ is comparable to $T_{sky}$ while at 100\,MHz it is about two orders of magnitude lower. The high noise at low frequency is due to  $\Re[Y_{self}]\rightarrow 0$ as the antenna becomes electrically very small. At 100\,MHz, the antenna is near resonance and $Y_{opt}$ becomes comparable to $Y_{self}$.

\subsection{Parallel Dipoles}
\label{sec:par_dip}
\begin{figure}[htb]
\centering
\noindent
  \includegraphics[width=2.75in]{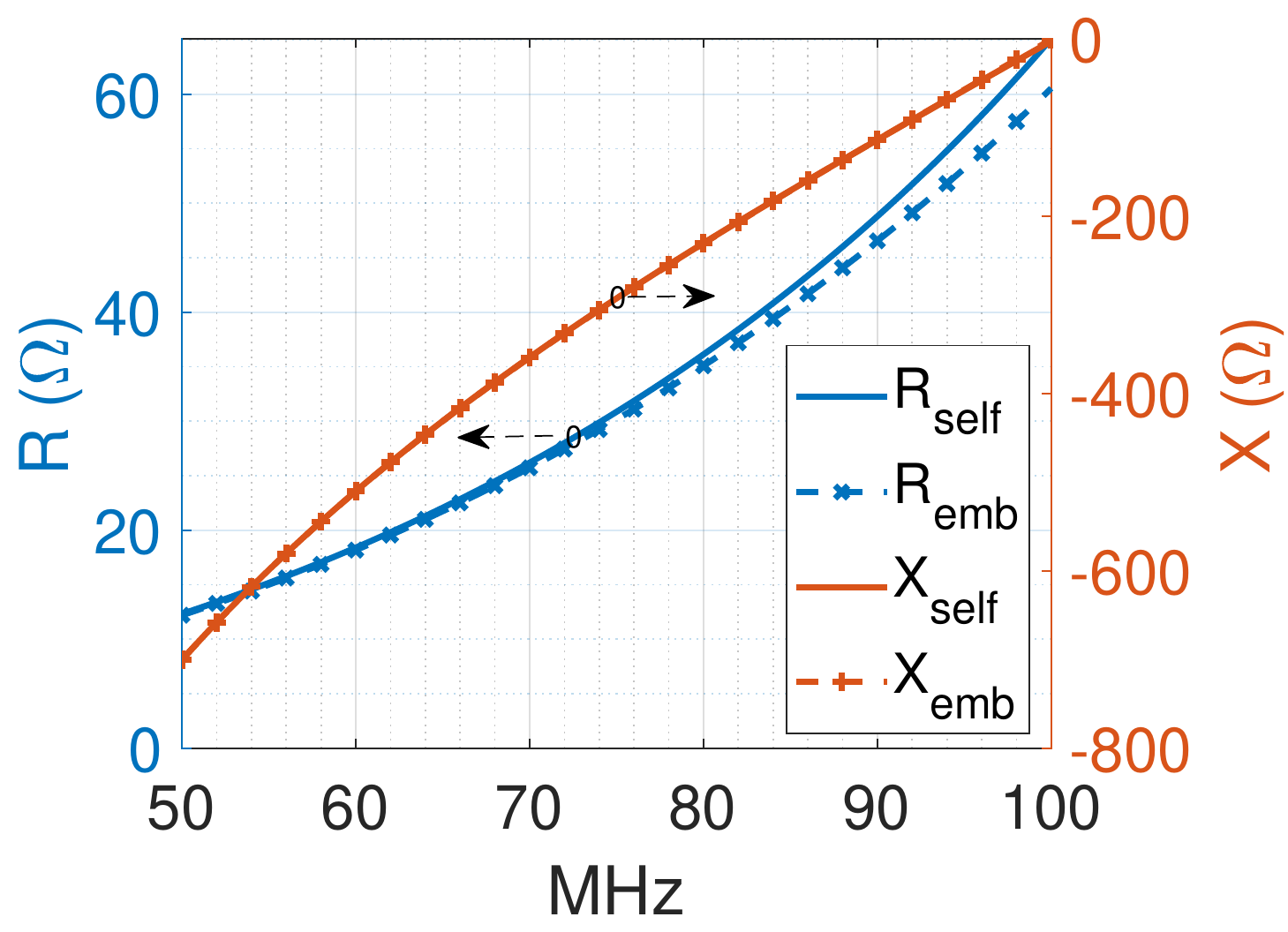}
\caption{Calculated self impedance of a thin dipole $L=\SI{1.44}{\metre}$ with wire radius of \SI{1}{\milli\metre}. The embedded impedance is computed for two identical parallel dipoles $d=\SI{0.9}{\metre}$ connected to SAV-541+ LNAs at $V_{DS}=\SI{3}{\volt}, I_{D}=\SI{60}{\milli\ampere}$ bias point.}
\label{fig:Z_self_embd_144}
\end{figure}

We compare the single dipole case with a two-element interferometer. The antenna length is kept at $L=\SI{1.44}{\metre}$. An identical dipole is placed parallel to the first at $d=\SI{0.9}{\metre}$ ($0.3\lambda$ at 100\,MHz). We connect the antennas to two SAV-541+ LNAs. Fig.~\ref{fig:Z_self_embd_144} shows the embedded impedance of the antenna which is not much different from the self impedance as expected because the LNA input impedance is high. 

\begin{figure}[htb]
\centering
\noindent
  \includegraphics[width=2.75in]{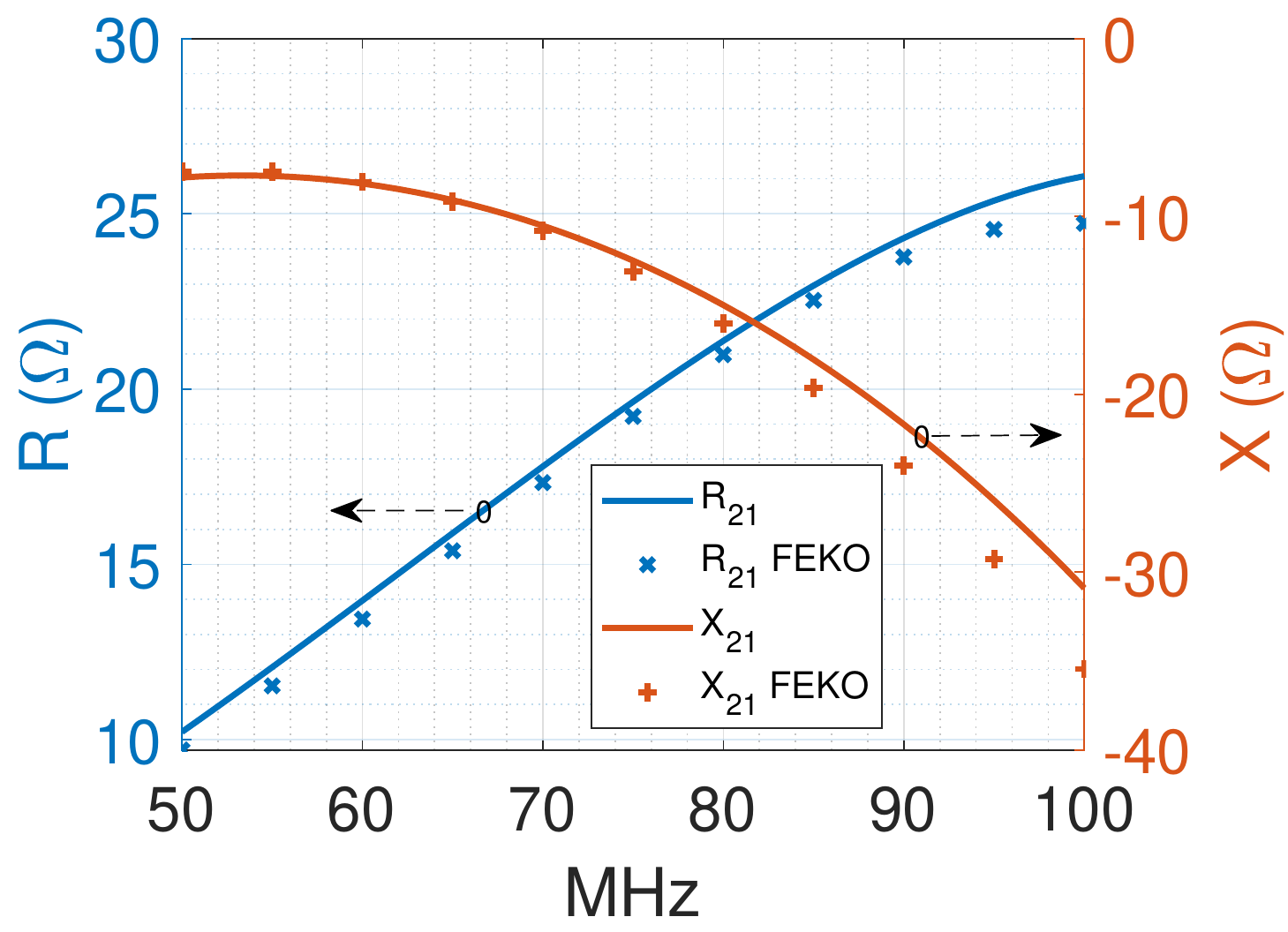}
\caption{Calculated and simulated mutual impedance of two identical parallel thin dipoles.}
\label{fig:Z21_L144_d09}
\end{figure}

The calculated mutual impedance is shown in Fig.~\ref{fig:Z21_L144_d09} which again is comparable to FEKO. Note that $R_{21}$ decreases with decreasing frequency because the antenna becomes electrically smaller. The self and mutual impedances are computed using the thin dipole formula in~\cite{Brown_1685626, Kraus_ch10}.

\begin{figure}[htb]
\centering
\noindent
  \includegraphics[width=3.25in]{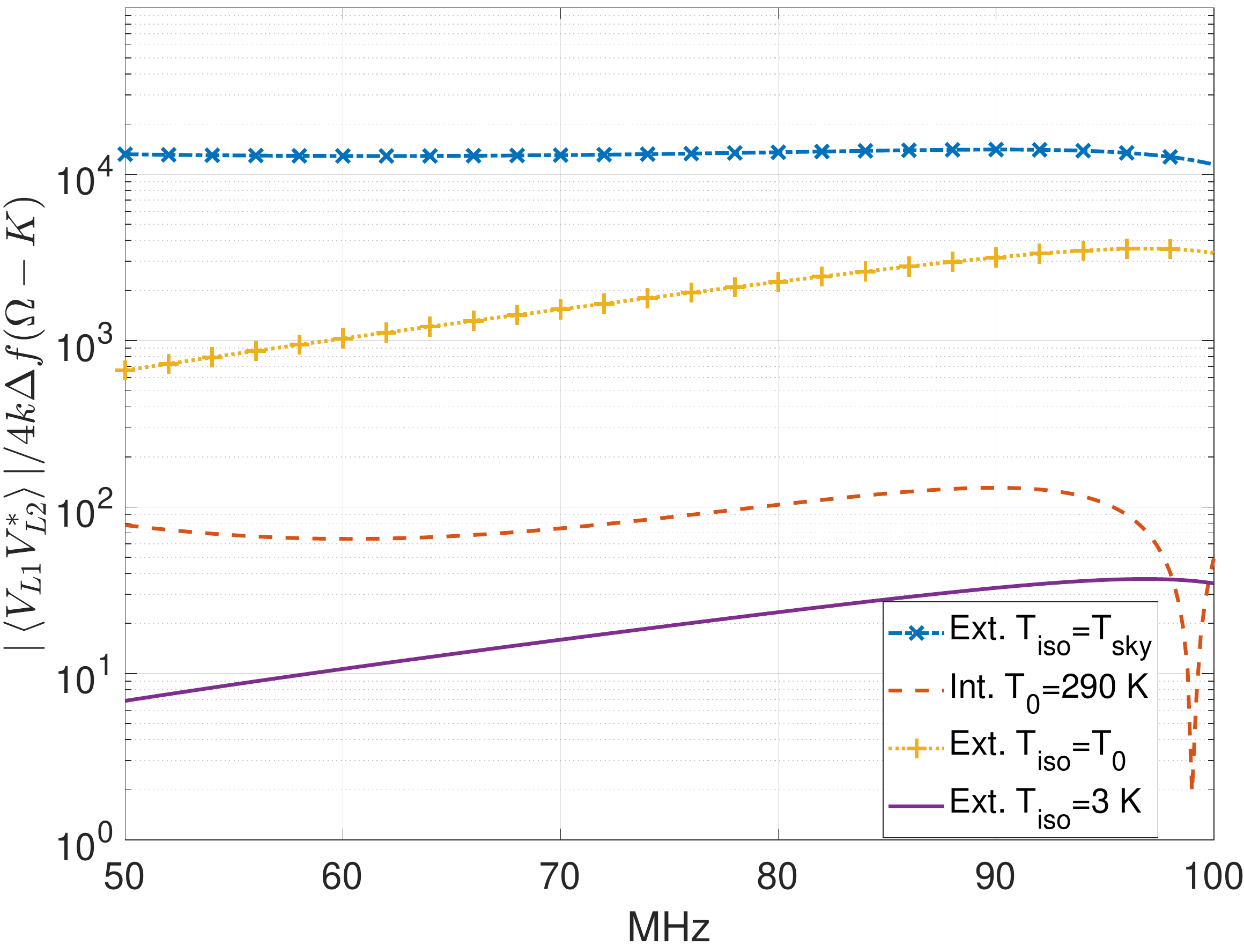}
\caption{Calculated mutual coherence (absolute value) due to internal and external noise of two identical parallel thin dipoles. $T_{iso}=T_0$ is the external noise temperature of an anechoic chamber while $T_{iso}=\SI{3}{\kelvin}$ represents the isotropic noise due to the CMB.}
\label{fig:ZVLVL_L144_d09}
\end{figure}

The mutual coherence due to external and internal noise sources computed using \eqref{eqn:ext_noise_corr_loaded} and \eqref{eqn:Int_noise_vL_tot} respectively, are reported in Fig.~\ref{fig:ZVLVL_L144_d09}. This quantity is normalized with respect to $4k\Delta f$ such that the unit is ohms-kelvin, \si{\ohm}-\si{\kelvin}. We see that the internal noise is consistently two orders of magnitude below signal due to $T_{iso}=T_{sky}$. This relative level is comparable to the minimum achievable noise temperature for the single element case in Fig.~\ref{fig:T_single}. Note that we achieve this performance with the two-element interferometer without any optimization effort.  

The mutual coherence due to internal noise sources is of course not zero. If the external noise is the \SI{3}{\kelvin} CMB, then the internal noise will be higher than the external signal by as much as an order of magnitude at 50\,MHz as shown in Fig.~\ref{fig:ZVLVL_L144_d09}. This means recovery of the desired signal requires precise knowledge of internal noise contribution, in particular the frequency variation thereof.

\subsection{Internal Noise Decorrelation}
\label{sec:noise Decorrelation}
Fig.~\ref{fig:ZVLVL_L144_d09} suggests that the mutual coherence due to internal noise falls in a null at approximately 99\,MHz which is possible as discussed earlier. This has been validated with a different approach given in~\cite{Warnick_5062509, warnick_maaskant_ivashina_davidson_jeffs_2018} using equations (5.51) and (5.52) in  \cite{warnick_maaskant_ivashina_davidson_jeffs_2018} and found to be in exact agreement (not shown).


On closer inspection, the zero crossing occurs at 99.04\,MHz where the factor $2\Re\left[Z_{21}(VV|_{V,I_{n2}})/(Z_{11}+Z_L)\right]$ in \eqref{eqn:Int_noise_vL_tot} crosses zero. In this instance, there is no simpler explanation. It occurs due to a more complex interaction than \eqref{eqn:zero_noise_coherence} suggests; this is fully expected because $Z_L, Z_{emb}, Z_{opt}$ are all capacitive such that the said condition is not fulfilled. However we can say that the internal noise decorrelation occurs close to the minimum of $|VV|_{V,I_{n2}}|$. The zero crossing can be moved beyond 100\,MHz by shortening the antenna to $L=\SI{1.3}{\metre}$ (at the cost of some reduction the external response at low frequencies). This suggests that the zero-crossing point is influenced by the series resonance frequency of the dipole. 

\section{Half-Space Systems}
\label{sec:half-space}
The two-dipole example is suspended in infinite space. We now consider more realistic systems that are deployable on the ground which occupies a semi-infinite half-space. There are two natural extensions to the two-dipole system: two monopoles and horizontal dipoles over ground. Ohmic losses in the soil emit noise proportional to the ambient temperature ($T_{amb}$) of the environment which is seen as partially correlated by the closely-spaced antennas. This is an important consideration, in particular for Cosmological signal detection. 

\begin{figure}[htb]
\centering
\noindent
  \includegraphics[width=2.5in]{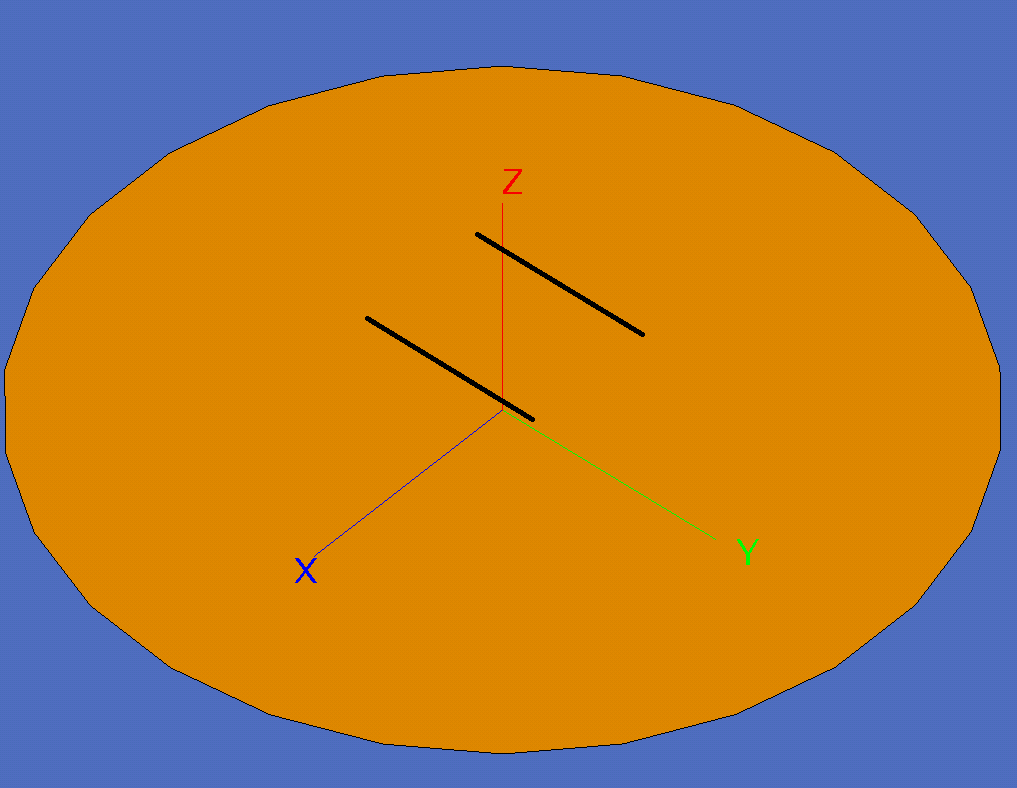}
\caption{Two closely-spaced dipoles over ground with the same parameters as Fig.~\ref{fig:dip_L13_d09_h07} simulated in FEKO. The ground plane is a circular perfect electric conductor (PEC) which for clarity is shown as  \SI{6}{\metre} in diameter; we also explore the effect of ground diameter to the ohmic loss noise contribution. The ground plane is situated over a soil model based on a Murchison Radio-astronomy Observatory (MRO) soil sample with 2\% humidity~\cite{7293140}.}
\label{fig:dip_over_gnd}
\end{figure}

We compared both the horizontal dipoles and monopoles over lossy ground with a finite conductive circular ground plane. As expected, the antenna radiation efficiency calculation results suggest that the monopole system is significantly more susceptible to soil noise than the dipole system for the same conductive ground plane size. For example, the radiation efficiency ($\eta_{rad}$) of the system in Fig.~\ref{fig:dip_over_gnd} is $\gtrsim95\%$ while the corresponding monopole system efficiency is 45\% to 60\% indicating the system is as sensitive to the soil as to the sky, which is not desirable. As a result, we focus on the analysis of the two horizontal dipole system in Fig.~\ref{fig:dip_over_gnd}.

\subsection{Dipoles Over Ground Screen}
\label{sec:dip_gnd}
It can be shown from image theory~\cite{Kraus_ch10} that the entries of the impedance matrix of the two dipoles over ground (subscript $._g$) where the dipole lengths are $\lesssim \lambda/2$ are well approximated by 
\begin{eqnarray}
Z_{11g}&\approx&Z_{11}-Z_{21:2h} \nonumber \\
Z_{21g}&\approx&Z_{21:d}-Z_{21:diag} 
\label{eqn:Zdip_gnd}
\end{eqnarray}
where $Z_{21:2h}, Z_{21:d}, Z_{21:diag}$ are the mutual impedance of two dipoles at distances $2h$ (the distance to self image of the dipole), $d$ (horizontal spacing) and $diag=\sqrt{d^2+(2h)^2}$ is the distance to the image of the opposing dipole, respectively. 

Equation \eqref{eqn:Zdip_gnd} implies that $R_{21g}\approx R_{21:d}-R_{21:diag}$. Referring back to Fig.~\ref{fig:R21_self} we see that $d=0.43\lambda$ demarcates the sign inversion of $R_{21}$ for $\sim\lambda/2$ dipole. This means that if $d=2h=0.4\lambda$ then $diag=0.57\lambda$ which suggest $R_{21g}>R_{21:d}$ because $R_{21:diag}$ is negative. However at $d=2h=0.2\lambda$, $diag=0.28\lambda$ therefore $R_{21g}<R_{21:d}$ as $R_{21:diag}$ is positive. This suggests that the dipole over ground system could be spaced slightly farther than the two dipoles in free space to take advantage of the sign inversion of $R_{21:diag}$.  

\begin{figure}[htb]
\centering
\noindent
  \includegraphics[width=2.75in]{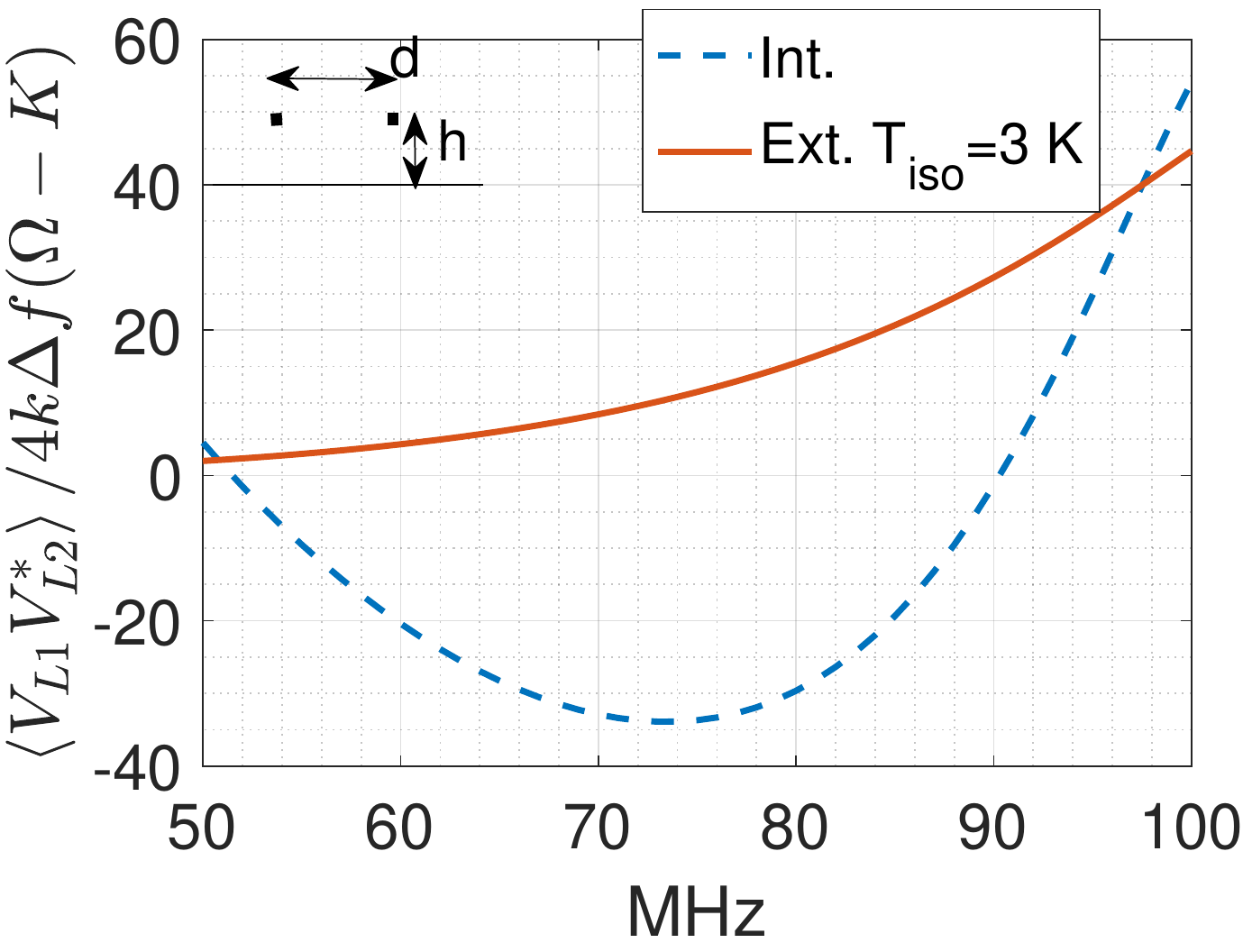}
\caption{Calculated mutual coherence due to internal and external noise of two identical parallel thin dipoles over infinite ground with length $L=\SI{1.3}{\metre}$, spacing $d=\SI{1}{\metre}$ and height over ground of $h=\SI{0.7}{\metre}$.}
\label{fig:dip_L13_d09_h07}
\end{figure}

Fig.~\ref{fig:dip_L13_d09_h07} reports the mutual coherence due to external and internal noise sources for $L=\SI{1.3}{\metre}$,  $d=\SI{1}{\metre}$ and $h=\SI{0.7}{\metre}$.  The internal noise again could be up to an order of magnitude higher than the external noise assuming thermal equilibrium at \SI{3}{\kelvin} and  the  curvature due to the internal noise sources is more significant than that due to the external noise. As expected, there is more difference between the response to the external noise at 50\,MHz and 100\,MHz compared to the monopole system. This is because of the sign inversion of $R_{21:diag}$ as discussed.  

\subsection{Noise From Ohmic Loss}
\label{sec:near}


To compute mutual coherence of noise due to ohmic losses, we revisit Sec.~\ref{sec:ext noise} to remove the assumption of thermal equilibrium between $T_{iso}$ and $T_{amb}$. The circuit model in Fig.~\ref{fig:antenna_noise} remains valid, however the external noise sources now comprise two components due to contributions of the far-field and ohmic loss.
\begin{eqnarray}
V_{e1}&=&V_{e1:ff}+V_{e1:\Omega} \nonumber \\
V_{e2}&=&V_{e2:ff}+V_{e2:\Omega} 
\label{eqn:ext_ohmic}
\end{eqnarray}
where the noise from ohmic loss is uncorrelated with that of the far-field sources. As for $V_{e1:ff}$ and $V_{e2:ff}$, \eqref{eqn:Vecorr} and \eqref{eqn:Vecorr_iso} still apply with a small modification of $\pi/2$ as the upper limit of $d\theta$ integration. The mutual coherence of the external noise voltages due to the ohmic losses, $V_{e1:\Omega}$ and $V_{e2:\Omega}$, is obtained by placing the antenna system in a test thermal equilibrium environment and taking the difference between the generalized Nyquist theorem and the far-field integration 
\begin{eqnarray}
\frac{\left<V_{e1:\Omega}V_{e2:\Omega}^*\right>}{4kT_{amb}\Delta f}&=& R_{12}-
\frac{\eta_0}{4\lambda^2}\int\limits_{0}^{2\pi}\int\limits_{0}^{\pi/2}  \bar{l}_1\cdot \bar{l}_2^*\sin\theta d\theta d\phi
\nonumber\\
&\triangleq&\Delta R_{12}
\label{eqn:ohmic_coherence}
\end{eqnarray}
and
\begin{eqnarray}
\frac{\left<|V_{e1:\Omega}|^2\right>}{4kT_{amb}\Delta f}&=& R_{11}-
\frac{\eta_0}{4\lambda^2}\int\limits_{0}^{2\pi}\int\limits_{0}^{\pi/2}  \norm{\bar{l}_1}^2 \sin\theta d\theta d\phi
\nonumber\\
&\triangleq&\Delta R_{11}
\label{eqn:ohmic_coherence_self}
\end{eqnarray}
With \emph{no} assumption regarding thermal equilibrium of external sources, we re-use \eqref{eqn:ext_noise_corr_loaded} with the following modification to \eqref{eqn:ext_noise_corr_loaded_B}
\begin{eqnarray}
&&\left<V_{th1}V_{th2}^*\right>|_{V_{e1,2}}= 
-2\Re\left[\frac{Z_{21}}{Z_{11}+Z_{L}}\right]\left< |V_{e1}|^2\right> \nonumber \\
&+&\left<V_{e1}V_{e2}^*\right>+\left<V_{e1}^*V_{e2}\right> \left(\left|\frac{Z_{21}}{Z_{11}+Z_{L}}\right|^2\right) 
\label{eqn:ext_noise_corr_loaded_wohmic} 
\end{eqnarray}
where
\begin{eqnarray}
\left< |V_{e1}|^2\right>&=&\left< |V_{e1:ff}|^2\right>+4kT_{amb}\Delta R_{11}\Delta f \nonumber\\ \left<V_{e1}V_{e2}^*\right>&=&\left<V_{e1:ff}V_{e2:ff}^*\right>+4kT_{amb}\Delta R_{12}\Delta f
\label{eqn:ext_noise_corr_loaded_Vqty} 
\end{eqnarray}
The error term due to uncalibrated ohmic loss is
\begin{eqnarray}
&&\frac{\Delta_\Omega\left<V_{th1}V_{th2}^*\right>|_{V_{e1,2}}}{4kT_{amb}\Delta f}= 
-2\Re\left[\frac{Z_{21}}{Z_{11}+Z_{L}}\right]\Delta R_{11} \nonumber \\
&+&\Delta R_{12}+\Delta R_{12}^* \left(\left|\frac{Z_{21}}{Z_{11}+Z_{L}}\right|^2\right) 
\label{eqn:error_ext_noise_corr_loaded_wohmic} 
\end{eqnarray}
This suggests that the dominant term is $\Delta R_{12}$ as expected. However, there is a contribution from $\Delta R_{11}$ which may not be negligible for highly inefficient antenna systems.

We compute the far-field integrals numerically using electric field samples at \SI{2}{\degree} resolution and study the effect of soil noise as a function of ground plane diameter from \SI{6}{\metre} to \SI{40}\metre. As expected, the ohmic noise contribution decreases with increasing ground plane diameter. At \SI{40}{\metre} diameter, the level of mutual coherence due to ohmic loss becomes comparable to the response of the system observing \SI{100}{\milli\kelvin} isotropic sky as shown in Fig.~\ref{fig:dipole_FEKO}.


\subsection{Cross-talk Consideration}
\label{sec:xtalk}
Cross-talk may be modeled as off-diagonal terms $g_{12}, g_{21}$ in $\mathbf{G}$ in \eqref{eqn:Rc}. The output of the correlator becomes
\begin{eqnarray}
\mathbf{C}_c(1,2)&=&g^*_{21}\left[g_{11}\left<|V_{L1}|^2\right>+g_{12}\left<V_{L1}^*V_{L2}\right>\right]\nonumber \\
&+&g_{22}^*\left[g_{11}\left<V_{L1}V_{L2}^*\right> +g_{12}\left<|V_{L2}|^2\right>\right]
\label{eqn:out_xtalk}
\end{eqnarray}
where $g_{11}, g_{22}$ are the diagonal entries of $\mathbf{G}$. The desired output in \eqref{eqn:out_xtalk} is $g_{22}^*g_{11}\left<V_{L1}V_{L2}^*\right>$ which is contaminated by additional terms. In theory if $\mathbf{G}$ is known, then its effect may be removed from the measurement. This calibration process is beyond our current scope. 

For now, we compute impact of \emph{uncalibrated} cross-talk. For this purpose, it suffices to let $g_{11}=g_{22}=1$, $g_{12}=g_{21}=x_{talk}$ and let the antennas be identical such that
\begin{eqnarray}
\mathbf{C}_c(1,2)
&=& \left<V_{L1}V_{L2}^*\right>+\left<V_{L1}^*V_{L2}\right>|x_{talk}|^2\nonumber \\
&+&2\Re[x_{talk}]\left<|V_{L1}|^2\right>
\label{eqn:out_xtalk_sim}
\end{eqnarray}
This suggests that the primary impact of cross-talk is \emph{the leakage of the single-element response}, $2\Re[x_{talk}]\left<|V_{L1}|^2\right>$, to the desired product. Assuming very closely space interferometer such that $\left<V_{L1}V_{L2}^*\right>\approx \left<|V_{L1}|^2\right>$, if the sky under observation is $\sim10^2$ to $\SI{e4}{\kelvin}$ and the desired error term is $\sim\SI{100}{\milli\kelvin}$, then the uncalibrated cross-talk term must be of the order \num{e-3} to \num{e-5} (\SI{-60}{\decibel} to \SI{-100}{\decibel}). This requires significant separation and shielding between the two branches of circuitry connected to the antennas such as the analog-to-digital (A/D) converters, printed circuit boards and power supplies.

Again with no assumption regarding thermal equilibrium of external sources, \eqref{eqn:ext_noise_corr_loaded_B} becomes
\begin{eqnarray}
&&\left<V_{th1}V_{th2}^*\right>|_{V_{e1,2}}= 
-2\Re\left[\frac{Z_{21}}{Z_{11}+Z_{L}}+x_{talk}\right]\left< |V_{e1}|^2\right> \nonumber \\
&+&\left<V_{e1}V_{e2}^*\right>+\left<V_{e1}^*V_{e2}\right> \left(\left|\frac{Z_{21}}{Z_{11}+Z_{L}}\right|^2+|x_{talk}|^2\right) 
\label{eqn:ext_noise_corr_loaded_xtalk} 
\end{eqnarray}
The error term due to uncalibrated $x_{talk}$ is 
\begin{eqnarray}
\Delta\left<V_{th1}V_{th2}^*\right>|_{V_{e1,2}}&=& 
-2\Re\left[x_{talk}\right]\left< |V_{e1}|^2\right> \nonumber\\ &+&\left<V_{e1}^*V_{e2}\right> |x_{talk}|^2
\label{eqn:err_ext_noise_corr_loaded_xtalk} 
\end{eqnarray}



\begin{figure}[htb]
\centering
\noindent
  \includegraphics[width=3.25in]{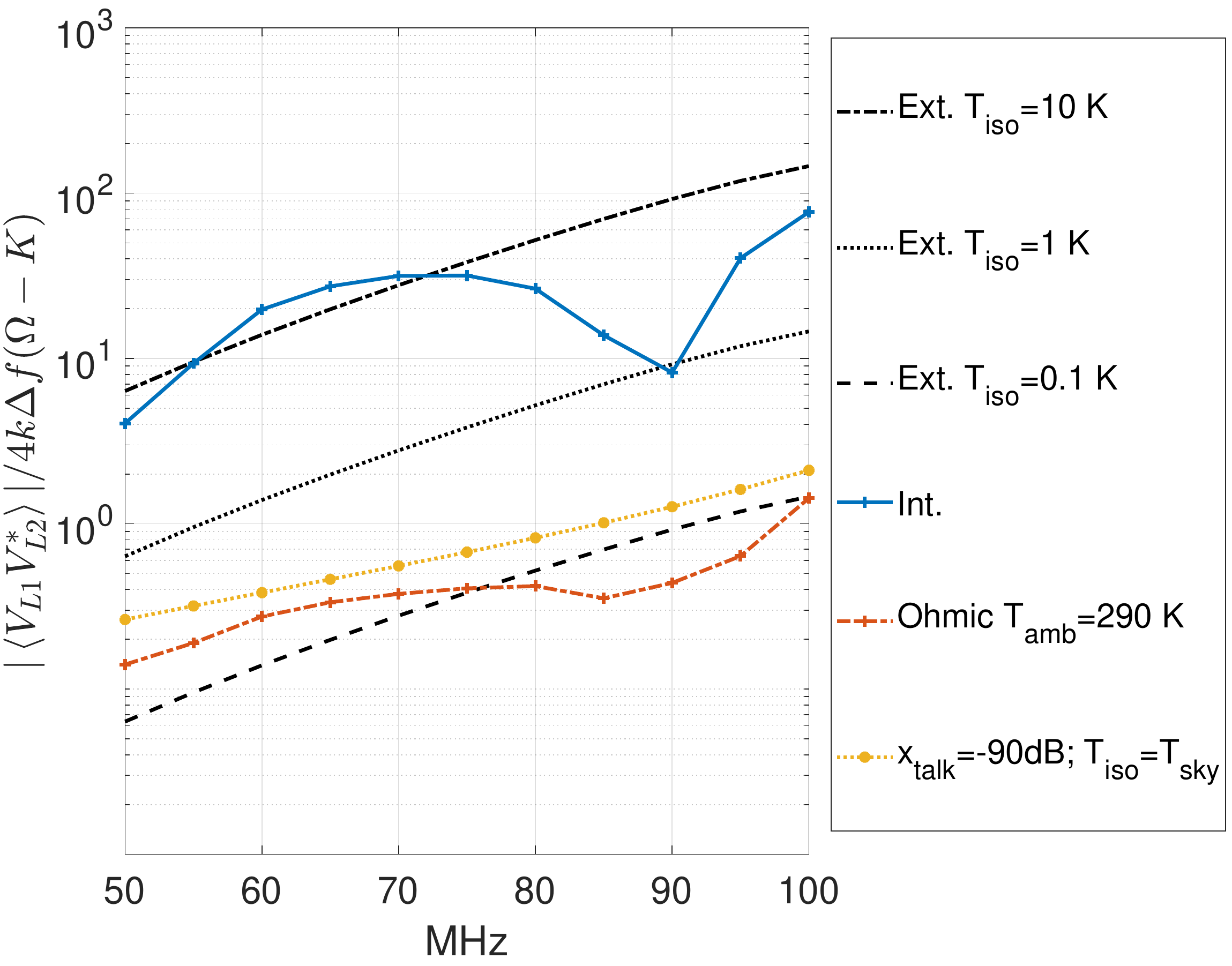}
\caption{Calculated mutual coherence (absolute value) due to internal noise, isotropic sky at 0.1 to \SI{10}{\kelvin}, ohmic loss and cross-talk assuming $x_{talk}=\SI{-90}{\decibel}$ for the two-dipole system with lengths $L=\SI{1.3}{\metre}$, spacing $d=\SI{1}{\metre}$ and height over ground of $h=\SI{0.7}{\metre}$. The PEC ground plane diameter is \SI{40}{\metre}.}
\label{fig:dipole_FEKO}
\end{figure}

Finally we illustrate the overall performance of the two dipole system using simulated data produced by FEKO. The internal noise is computed using \eqref{eqn:Int_noise_vL_tot} and \eqref{eqn:plug_noise_par_simp}. The noise due to ohmic losses is calculated using \eqref{eqn:error_ext_noise_corr_loaded_wohmic}  with $T_{amb}=\SI{290}{\kelvin}$. The antennas system is placed under isotropic sky as per \eqref{eqn:Vecorr_iso} with $T_{sky}=60\lambda^{2.55}\,\si{\kelvin}$. The calculated results are shown in Fig.~\ref{fig:dipole_FEKO}. The LNAs are the same Minicircuits part as in Sec.~\ref{sec:calc}. The constant cross-talk level of \SI{-90}{\decibel} is assumed over frequency which produces mutual coherence at a level comparable to that of an isotropic sky at a few hundred \si{\milli\kelvin}. We see that the contribution of internal noise of up to approximately \SI{10}{\kelvin} is the highest contributor to systematic error. It suggests that calibration of internal noise coupling is required for close-spacing interferometry for Cosmological signal detection.



\section{Conclusion}
\label{sec:conclusion}
We have examined the closely-spaced interferometer system through design formulas and an equivalent circuit. Our results show that it is possible to design a closely-spaced two-element interferometer with spectrally smooth response that is sensitive to the highly diffuse noise sources over 2:1 bandwidth (50-100\,MHz) desirable in Cosmological signal detection. However, mutual coherence due to internal noise coupling does not vanish as is commonly assumed. On the contrary, noise coupling is an inherent property of a closely-spaced interferometer as the mutual resistance that produces the zero-spacing response also couples the internal noise. Using the design formulas, we have also explored the contributions due to ohmic loss and cross-talk. The former may be mitigated with a highly conductive large ground ($\sim$  tens of \si{\metre} in dia.) and the later requires shielding and separation of the circuitry connected to the antennas.


\end{document}